\documentclass[prl,aps,
showpacs,
twocolumn,
nofootinbib,
floatfix,
superscriptaddress]{revtex4}
\usepackage{graphicx}
\usepackage{epsfig}
\usepackage{rotating}
\usepackage{amssymb}
\usepackage{dsfont}
\usepackage{psfrag}
\usepackage{amsmath,euscript,array,mathrsfs}

\newcommand{\SP}[1]{\begin{equation}\begin{split} #1
\end{split}\end{equation}}

\newcommand{\beq}{\begin{equation}}
\newcommand{\eeq}{\end{equation}}
\newcommand{\beqs}{\begin{eqnarray}}
\newcommand{\eeqs}{\end{eqnarray}}

\def\hbar{\hspace{0pt}\raisebox{1pt}{$-$} \hspace{-7pt} h}

\def\di{\mbox{d}}
\def\r{\rho}

\newcommand{\be}{\begin{equation}}
\newcommand{\ee}{\end{equation}}
\newcommand{\bea}{\begin{eqnarray}}
\newcommand{\eea}{\end{eqnarray}}

\def\lbldef#1#2{\expandafter\gdef\csname #1\endcsname {#2}}

\def\href#1#2{#2}

\newcommand{\ber}{\begin{eqnarray}}
\newcommand{\eer}{\end{eqnarray}}

\newcommand{\beqar}{\begin{eqnarray}}

\newcommand{\eeqar}{\end{eqnarray}}

\newcommand{\dsl}
  {\kern.06em\hbox{\raise.15ex\hbox{$/$}\kern-.56em\hbox{$\partial$}}}

\newcommand{\eeqarr}{\end{eqnarray}}
\newcommand{\ZZ}{{\rm \kern 0.275em Z \kern -0.92em Z}\;}

\def\CC{{\mathchoice
{\rm C\mkern-8mu\vrule height1.45ex depth-.05ex
width.05em\mkern9mu\kern-.05em}
{\rm C\mkern-8mu\vrule height1.45ex depth-.05ex
width.05em\mkern9mu\kern-.05em}
{\rm C\mkern-8mu\vrule height1ex depth-.07ex
width.035em\mkern9mu\kern-.035em}
{\rm C\mkern-8mu\vrule height.65ex depth-.1ex
width.025em\mkern8mu\kern-.025em}}}

\def\RR{{\rm I\kern-1.6pt {\rm R}}}

\def\ZZ{{\rm Z}\kern-3.8pt {\rm Z} \kern2pt}
\def\IB{\relax{\rm I\kern-.18em B}}
\def\ID{\relax{\rm I\kern-.18em D}}
\def\II{\relax{\rm I\kern-.18em I}}
\def\IP{\relax{\rm I\kern-.18em P}}

\newcommand{\bear}{\begin{eqnarray}}
\newcommand{\eear}{\end{eqnarray}}

  \def\w{\omega}
\def\r{\rho}                                     

\def\6{\partial}

\def\bea{\begin{eqnarray}}
\def\eea{\end{eqnarray}}

\def\beqx{\begin{displaymath}}
\def\eeqx{\end{displaymath}}

\newcommand{\bmat}{\left(\begin{array}}
\newcommand{\emat}{\end{array}\right)}

\def\r{\rho}

\def\bo{{\raise-.3ex\hbox{\large$\Box$}}}               
\def\face{{\raise.2ex\hbox{$\displaystyle \bigodot$}\mskip-2.2mu \llap {$\ddot
        \smile$}}}                                   
\def\>{\rangle}                                      
\def\<{\langle}                                      

\def\leftrightarrowfill{$\mathsurround=0pt \mathord\leftarrow \mkern-6mu
        \cleaders\hbox{$\mkern-2mu \mathord- \mkern-2mu$}\hfill
        \mkern-6mu \mathord\rightarrow$}        
\def\dvec#1{\vbox{\ialign{##\crcr
        \leftrightarrowfill\crcr\noalign{\kern-1pt\nointerlineskip}
        $\hfil\displaystyle{#1}\hfil$\crcr}}}           

\def\-{\hphantom{-}}

\newcommand{\wedg}{_{\wedge}}

\begin{document}
\title{A light scalar from walking solutions in gauge-string duality.
}

\author{Daniel Elander, Carlos N\'u\~nez and Maurizio Piai}
\affiliation{Swansea University, School of Physical Sciences,
Singleton Park, Swansea, Wales, UK}

\date{\today}

\begin{abstract}
We consider the  type-IIB background generated 
by  the strong-coupling limit  of $N_c$ $D5$ branes wrapped on  $S^2$, 
and focus our attention on a special class of solutions that exhibit {\it walking} behavior.
We compute numerically the spectrum of scalar fluctuations around vacua of this class.
Besides two cuts, and sequences of single poles converging on one of the branch points,
the spectrum contains one isolated scalar, the mass of which is suppressed by the 
length of the walking region.
Approximate scale-invariance symmetry
in the walking region suggests that this be interpreted
 as a light dilaton, the pseudo-Goldstone boson of dilatations.
\end{abstract}


\maketitle

\section{introduction}

Theories with strongly-coupled approximate infrared fixed points,
such as for instance~\cite{WTC},  
are difficult to study. The notion itself of {\it approximate}
scale-invariance, which they imply,  
is  very elusive, and its dynamical implications obscure.  
In particular, it is an open question
whether  such a dynamical feature requires the existence of a light
scalar in the low-energy spectrum, the {\it dilaton},
remnant of the spontaneous breaking of scale invariance.
 The consequences of its existence  
would be very dramatic in phenomenological applications, 
including dynamical electro-weak symmetry breaking~\cite{dilaton}.

In this paper, we address this specific question 
in the context of  gauge-gravity correspondence.
We focus on one specific set-up, based on a 10-dimensional type-IIB
string theory background, generated by taking the strong-coupling limit
of the system~\cite{MN} of $N_c$ stacked $D5$ branes, wrapped on an $S^2$
in an internal CY3 manifold.
A large class of solutions
to this system (with no flavor degrees of freedom, $N_f=0$) 
behaves in the IR in a way very similar to 
what is expected in the presence of an approximate fixed point~\cite{NPP}.
Namely, a suitably defined gauge coupling~\cite{VLM}
is finite and approximately constant
over a sizable intermediate-energy regime ({\it walking}).

We carry out the study of the scalar perturbations
to this class of solutions, and study numerically the discrete mass spectrum.
In order to do so, we draw heavily on~\cite{HNP} and~\cite{BHM},
and apply the formalisms developed there to 
the new specific solutions we are interested in. In particular, it was found in~\cite{BHM} that the 10d system admits a consistent truncation to a 5d non-linear sigma model consisting of six scalar fields coupled to gravity. Furthermore, a formalism was developed (and generalized in~\cite{Elander:2009bm}) in which it is possible to study the fluctuations of these six scalar fields as well as of the 5d metric (also considered dynamical), and where equations of the scalar fluctuations effectively decouple from those of the 5d-gravity degrees of freedom. This renders the present study technically feasible.

\section{Wrapped D5 system}
We start from the action of type-IIB supergravity,
truncated to include only the graviton, dilaton and  RR 3-form $F_3$ of 
flux proportional to $N_c$ (the number of colors of the dual field 
theory),
\beq
S_{IIB}=\frac{1}{G_{10}}\int d^{10}x \sqrt{-g}\left[
R-\frac{1}{2}(\partial\phi)^2
-\frac{e^{\phi}}{12}F_3^2  \right] 
\,.
\label{actioniibsources}\eeq
We propose a background solution
where the functions appearing in it depend 
only on the radial coordinate $\r$, 
and not on the Minkowski coordinates $x^\mu$ nor on the 5 angles 
$(\theta,\tilde{\theta},
\phi,\tilde{\phi},\psi)$:
\begin{widetext}
\bea
&\frac{d s_E^2}{\mu^2}= e^{ 2 f } \Big[\frac{dx_{1,3}^2}{\mu^2} +
e^{2k}d\rho^2
+ e^{2 h}
(d\theta^2 + \sin^2\theta d\varphi^2) \nonumber
+\frac{e^{2 \hat{g}}}{4}
\left((\tilde{\omega}_1+a d\theta)^2
+ (\tilde{\omega}_2-a\sin\theta d\varphi)^2\right)
 + \frac{e^{2 k}}{4}
(\tilde{\omega}_3 + \cos\theta d\varphi)^2\Big], \nonumber\\
&\frac{F_{3}}{\mu^2}=\frac{-N_c}{4}\Big[(\tilde{\omega}_1+b d\theta)\wedg
(\tilde{\omega}_2-b \sin\theta d\varphi) \wedg
(\tilde{\omega}_3 + \cos\theta d\varphi)+
 b'd\rho \wedg (d\theta \wedg \tilde{\omega}_1  -
\sin\theta d\varphi
\wedg
\tilde{\omega}_2) - (1-b^2) \sin\theta d\theta\wedg d\varphi \wedg
\tilde{\omega}_3\Big].
\label{nonabmetric424}
\eea
\end{widetext}

We have defined $\mu^2=\alpha' g_s$, and used the $SU(2)$ left-invariant one forms,
$
\tilde{\w}_1= \cos\psi d\tilde\theta\,+\,\sin\psi\sin\tilde\theta
d\tilde\varphi$, 
$\tilde{\w}_2=-\sin\psi d\tilde\theta\,+\,\cos\psi\sin\tilde\theta
d\tilde\varphi$ and  
$\tilde{\w}_3=d\psi\,+\,\cos\tilde\theta d\tilde\varphi$.
Following~\cite{HNP}, and fixing some integration constants, the
 background can be  rewritten as $\phi= 4f$, and
\bea
&& 4 e^{2h}=\frac{P^2-Q^2}{P\coth(2\rho) -Q}, \;\; e^{2\hat{g}}= 
P\coth(2\rho) 
-Q,\;\;\nonumber\\
&&e^{2k}= \frac{P'}{2}\,,\;\;
 \sinh(2\rho)a=\frac{P}{(P\coth(2\rho) 
-Q)},\;\; \nonumber\\
&&b=\frac{(2N_c)\rho}{\sinh(2\rho)}\,,\;\; e^{4\phi-4\phi_0}= \frac{8 
\sinh^2(2\rho)}{(P^2-Q^2)P'}\,,
\label{functions}
\eea
where
\bea
 Q(\rho)=N_c (2\rho \coth(2\rho) -1)
\label{BPSeqs}
\eea
and $P(\rho)$ satisfies a
 decoupled second order equation 
\beq
P'' + P' \Big(\frac{P'+Q'}{P-Q} +\frac{P'-Q'}{P+Q} - 4 
\coth(2\rho)
\Big)=0\,.
\label{master}
\eeq

\subsection{Walking solutions}

We are looking for solutions to Eq.~(\ref{master})
which in the UV are just tiny perturbations
of the special $\hat{P}\equiv 2N_c\r$  solution~\cite{MN}, but that become 
approximately constant below some finite $\r_{\ast}$, analogous to 
the solutions in~\cite{NPP}.
We can do this by linearizing a perturbation around $\hat{P}$,
 by assuming that the solution can be written as
$
P(\r)=\hat{P}(\r)\,+\,\varepsilon p(\r)$\,,
and replacing in Eq.~(\ref{master}).
For large-$\r$ ,
the resulting equation can be solved exactly, and neglecting power-law corrections, 
the solution
behaves as
\beqs
p(\r)&\sim&c_1 e^{-4 \r} \,+c_2 e^{2\r}\,,
\eeqs
implying that consistency of the perturbative expansion towards $\r \rightarrow \infty$ enforces the choice $c_2=0$.

\begin{figure}[h]
\includegraphics[width=7cm]{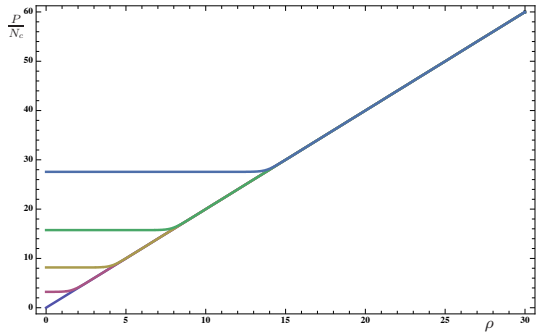}
\caption{Examples of solutions to Eq.~(\ref{master}) which fall into the
class with {\it walking} behavior, compared to $\hat{P}=2N_c \r$.}
\label{Fig:solutions}
\end{figure}

We can use this result  in setting up the boundary conditions 
(at large-$\r$) and numerically solve Eq.~(\ref{master}) toward the IR. 
By inspection, these solutions are precisely the ones we 
were looking for: they start deviating significantly from $\hat{P}$ 
at some $\r_{\ast}>0$, below which they become approximately constant.
We plot a few examples of such solutions in Fig.~\ref{Fig:solutions}.

\begin{figure}[h]
\includegraphics[width=7cm]{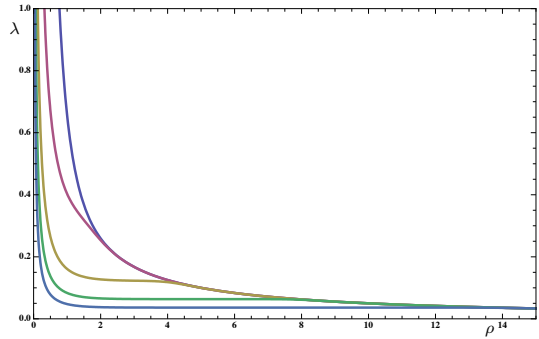}
\caption{The gauge coupling $\lambda$ defined in the text,
as a function of the radial coordinate, for the same
solutions as in Fig.~\ref{Fig:solutions}.}
\label{Fig:running}
\end{figure}

Following~\cite{VLM}, we define the four-dimensional gauge coupling to be:
\beqs
\lambda\,=\,\frac{g^2 N_c}{8\pi^2}&\equiv&\frac{N_c \coth\r}{P}\,.
\label{Eq:gauge}
\eeqs
We plot in Fig.~\ref{Fig:running} the results obtained for the $\hat{P}$ solution and
the same sample of solutions as in Fig.~\ref{Fig:solutions}.
Notice the three distinct behaviors. Near the IR ($\r\rightarrow 0$), 
the running leads to a divergence. Near the UV ($\r\rightarrow \infty$), 
the coupling is vanishing. In the intermediate region $\r_{I}<\r<\r_{\ast}$,
where $\r_{I}\sim {\cal O}(1)$,
the gauge coupling is effectively constant ({\it walking}).

\section{Five-dimensional supergravity formalism.}
Following~\cite{BHM}, we  define new variables
$[A,g,p,x,\phi,a,b]$ as
\bea
& & f = A + p - \frac{x}{2}, \;\;
\tilde g = -A - \frac{g}{2} - p + x + \log 2, \nonumber\\
& & h = -A + \frac{g}{2} - p + x, \;\;\; k = -A - 4 p + \log 2, 
\eea
replace in the background equations from Eq.~(\ref{actioniibsources}) 
and change the radial coordinate according to
$
d\r e^{A+k}= dz
$\,.
The resulting system admits a consistent truncation to an effective 5-dimensional non-linear sigma model with fields 
$\Phi^i=[g,p,x,\phi,a,b]$ coupled to gravity: 
\beq
{\cal L}_{5d}^{eff}=\frac{4\mu^4 N_c^2(4\pi)^3}{G_{10}}
\sqrt{g}\left[\frac{R}{4}-\frac{1}{2}G_{ab}\partial \Phi^a \partial \Phi^b 
-V(\vec{\Phi})\right] \,,
\label{5deffnonab}
\eeq
in a  space-time of the form
\beq
ds^2= e^{2A}dx_{1,3}^2 + dz^2\,.
\eeq

The sigma-model metric is
\bea
\label{sigmamodelmetric}
 4G_{\phi\phi}= 2 G_{gg}= 
G_{xx}=\frac{G_{pp}}{6}=1\,,\\
\nonumber G_{aa}=\frac{e^{-2g}}{2},\;\; G_{bb}= 
\frac{N_c^2 e^{\phi-2x}}{32
},
\label{kahler}
\eea
and the potential $V$ is given by (in agreement with~\cite{BHM})
\begin{widetext}
\SP{
	V = \frac{e^{-2 (g+2 (p+x))}}{128} \Bigg[& 16 \left(a^4+2
   \left(\left(e^g-e^{6 p+2 x}\right)^2-1\right) a^2+e^{4 g}-4
   e^{g+6 p+2 x} \left(1+e^{2 g}\right)+1\right)+ \\& e^{12 p+2 x+\phi
   } \left(2 e^{2 g} (a-b)^2+e^{4 g}+\left(a^2-2 b
   a+1\right)^2\right) N_c^2 \Bigg].
}
\end{widetext}

\section{Fluctuations,  boundary conditions and numerical study}

From here on, we work in units where $\mu^2 N_c=1$.
Following~\cite{BHM}, after changing coordinates 
$dz = e^{A+k} d\r$,  writing the fluctuations as $\mathfrak{a}(x,\r)=e^{i K x}\mathfrak{a}(\r)$, 
replacing $\Box$ by $-K^2$ (=$M^2$) and using the equations
for the 5d gravity fluctuations,  
 we obtain the system of linearized equations for the scalar perturbations \cite{Elander:2009bm}
\SP{
\label{eq:eomfluc}
	\Big[ D_z^2 + 4 A' D_z - e^{-2A} K^2 \Big] \mathfrak{a}^a - \Big[ V^a_{|c} - \mathcal{R}^a_{\ bcd} \Phi'^b \Phi'^d + &\\ \frac{4 (\Phi'^a V_c + V^a \Phi'_c )}{3 A'} + \frac{16 V \Phi'^a \Phi'_c}{9 A'^2} \Big] \mathfrak{a}^c = 0&.
}
It is only for particular values of $K^2$ that the coupled differential equations \eqref{eq:eomfluc} allow 
for solutions with acceptable IR and UV behavior. These $K^2 = -M^2$  give us the glueball spectrum of the dual field theory. 
In order to find them, we  employ a numerical method described in \cite{BHM}, that 
in effect evolves solutions from both the IR and the UV, and then determines whether they  match  smoothly at a midpoint.

In the UV (for $\r\rightarrow +\infty$), Eq.~(\ref{eq:eomfluc})
can be diagonalized by a change to a  basis  
in which the fluctuations behave as
$\psi^i = e^{C_i \rho} \sum_n a_{i,n} \rho^{b_{i,n}}$,
where the exponents $b_{i,n}$ in general are non-integer, while
\SP{
    C_{1,2,6} = -1 \pm \sqrt{9 - M^2}, \\
    C_{3,4,5} = -1 \pm \sqrt{1 - M^2}.
}
Notice that this behavior implies the presence of cuts in the two-point functions
for $M^2>1$ and $M^2>9$.
We are interested here in the discrete spectrum, hence in the subleading behavior.
We choose  the minus signs in these expressions
and use the result to set up the UV boundary conditions. 
Note that the consistent truncation used in \cite{BHM}, when studying the perturbations
around $\hat{P}$, corresponds to keeping 
only the fluctuations that fall off as $e^{(-1 - \sqrt{1 - M^2}) \rho}$ in the UV.

The following expansion of $P$ holds in the IR~\cite{HNP}:

\bea
    P &=& P_0
    +\frac{4}{3} c_+^3 P_0^2 \rho^3
   +\frac{16}{15} P_0^2 c_+^3 \rho^5
   +\mathcal{O}(\rho^6)\,,
\eea
where $P_0$ and $c_+$ are integration constants.  
Requiring that the background asymptotes to $\hat{P}$ 
in the UV makes $c_+$ a function of $P_0$ (the analytical form of which is unknown).

We expand the fluctuations for $\rho\rightarrow 0$ as
$
    \mathfrak{a}^a = \sum_{n=-\infty}^\infty \mathfrak{a}^a_n \rho^n,
$
and plug them into the differential equation. The solutions are determined by the 12 integration constants
$\mathfrak{a}^1_0$, $\mathfrak{a}^1_1$, $\mathfrak{a}^2_0$, $\mathfrak{a}^2_1$, $\mathfrak{a}^3_0$, $\mathfrak{a}^3_1$, $\mathfrak{a}^4_0$, $\mathfrak{a}^4_1$, $\mathfrak{a}^6_{-1}$, $\mathfrak{a}^6_2$, $\mathfrak{a}^5_0$, and $\mathfrak{a}^5_3$,
as
\SP{
\begin{cases}
	\mathfrak{a}^1 = \mathfrak{a}^1_0 + \mathfrak{a}^1_1 \rho + 4 (\mathfrak{a}^5_0 - \mathfrak{a}^1_0) \rho^2 + \mathcal O \left( \rho^3 \right), \\
	\mathfrak{a}^2 = \mathfrak{a}^2_0 + \mathfrak{a}^2_1 \rho + \mathcal O \left( \rho^3 \right), \\
	\mathfrak{a}^3 = \mathfrak{a}^3_0 + \mathfrak{a}^3_1 \rho + \mathcal O \left( \rho^3 \right), \\
	\mathfrak{a}^4 = \mathfrak{a}^4_0 + \mathfrak{a}^4_1 \rho + \mathcal O \left( \rho^3 \right), \\
	\mathfrak{a}^5 = \mathfrak{a}^5_0 + (-4 \mathfrak{a}^1_0 + 2 \mathfrak{a}^5_0) \rho^2 + \mathfrak{a}^5_3 \rho^3 + \mathcal O \left( \rho^4 \right), \\
	\mathfrak{a}^6 = \mathfrak{a}^6_{-1} \rho^{-1} -\frac{2}{3} \mathfrak{a}^6_{-1} \rho + \mathfrak{a}^6_2 \rho^2 + \mathcal O \left( \rho^3 \right),
\end{cases}
}
Regularity in the IR requires (taking~\eqref{sigmamodelmetric} into account)
$\mathfrak{a}^6_{-1} = 0$ and $\mathfrak{a}^5_0=0$.
We use Dirichlet boundary conditions 
for the first four fields.

 \begin{figure}[htpb]
\includegraphics[width=7cm]{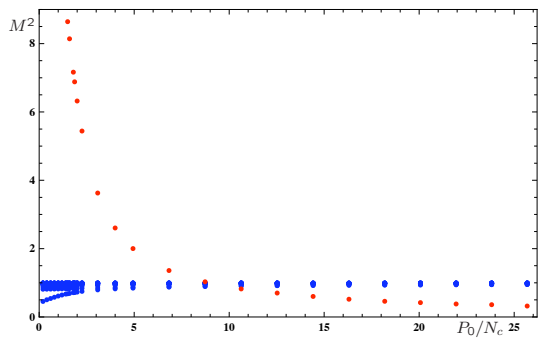}
\caption{The masses $M^2$ plotted against   $P_0/N_c \simeq 2\r_{\ast}$.
\label{Fig:walking}}
\end{figure}

 The numerical results are plotted in Fig.~\ref{Fig:walking}. 
 The spectrum for $\hat{P}$ ($P_0\rightarrow 0$) consists of a series of poles approaching $M^2=1$ (in agreement with~\cite{BHM}).
For large values of $P_0$ (equivalently, of $\r_{\ast}$), all the masses in the series 
approach the branch points, so that  the discrete spectrum effectively disappears into the
continuum. With one notable exception: one of the states
becomes lighter, and as $\r_{\ast}$ is increased, its mass is pushed below both the continuum thresholds. For $\r_{\ast}\rightarrow \infty$, this state becomes massless.

\section{Symmetry considerations}

Expressing the five-dimensional background metric $\di s^2=e^{2A}\left(\di x_{1,3}^2+e^{2k}\di \r^2\right)$ 
in terms of  $P$ and $Q$ yields
\beqs
e^{2A}&=&
\left(\frac{e^{\phi_0}}{4}\right)^{4/3}\,\left(
\sinh^2(2\r)\left(P^2-Q^2\right)\right)^{1/3}\,.
\eeqs
For  $P_0\gg 1$, there exists a region $\r_{I}\ll \r \ll \r_{\ast}$
where 
\beqs
P&\simeq&P_0\,\gg Q\,\simeq 2N_c \r\,,\\
\sinh^2 2\r&\simeq&\frac{e^{4\r}}{4}\,,\\
P^{\prime}&\simeq&\frac{c_+^3P_0^2}{4}e^{4\r}\,,
\eeqs
and in this region the metric is approximated by
\beqs
\di s^2&\simeq&\frac{e^{\frac{4\phi_0}{3}}P_0^{2/3}}{4^{5/3}}
\left(e^{\frac{4\r}{3}}\di x^2+\frac{c_+^3P_0^2e^{\frac{16\r}{3}}}{8}\di \r^2\right)\,.
\eeqs
Defining the scaling transformation
\beqs
\r&\rightarrow&\r+\Lambda\,,\\
x&\rightarrow&e^{{2\Lambda}{}}x\,,
\eeqs
the metric is conformal to itself: $\di s^2\rightarrow e^{16\Lambda/3}\di s^2$.

This  suggests that the light scalar might be  a light dilaton, 
and  hence its couplings should be dictated by 
this property.
With present  information, we are not able to reconstruct 
its composition in terms of the 
original degrees of freedom in the sigma-model. The metric is
not asymptotically AdS  in the UV, hence the rigorous procedure for holographic renormalization is not known,
nor is it  known how to characterize this state as normalizable 
or non-normalizable. A more detailed study of this and related points will be presented elsewhere~\cite{ENP}.

\section{Comments}

This is not a walking technicolor theory, since it does not yield a mechanism 
for electro-weak symmetry breaking.
However, the set of results collected here supports the idea that this system is a very
interesting laboratory, in which walking  can be studied systematically,
and in which dynamical questions can be addressed in a calculable form,
providing a guidance for  model building.

The class of solutions we found yields the four-dimensional gauge coupling 
of a walking theory (the Lagrangian of which, for present purposes, we do not need to know), in the sense that there is an
intermediate region $\r_{I}<\r<\r_{\ast}$ 
where the gauge coupling is approximately constant. While the interpretation in terms of the dual field theory is at this point not well understood, the very fact that we observe a particle in the spectrum with a mass much lower than the dynamical scale of the theory (the main result of this paper) suggests that its existence is due to the spontaneous breaking of an approximate symmetry. If this symmetry is scale invariance, as suggested in the previous section, then the light scalar would be interpreted as the dilaton, the pseudo-Goldstone boson of dilatations. From the gravity point of view, it is clear that scale invariance is broken in the IR by the gaugino condensate, and in the UV at the scale set by $\r_{\ast}$.

\vspace{1.0cm}
\begin{acknowledgments}
We thank A.~Armoni, S.~P.~Kumar, and I.~Papadimitriou for useful discussions.
The work of MP is supported in part by WIMCS. The work of DE is supported in part by STFC Doctoral Training Grant ST/F00706X/1.

\end{acknowledgments}


\end{document}